\documentclass[fleqn,twoside]{article}
\usepackage{latexsym} 
\usepackage{espcrc2}

\newcommand{\AmS}{{\protect\the\textfont2 A\kern-.1667em\lower.5ex\hbox{M}\kern-.125emS}}
\title{Generalized parity transformations in lattice Chern-Simons
  theory}

\author{C.D.~Fosco\address[CAB]{Centro At{\'o}mico Bariloche and
    Instituto Balseiro,
    Comisi{\'o}n Nacional de Energ{\'\i}a At{\'o}mica,\\
    8400 S.C. de Bariloche, Argentina}\thanks{Supported by CONICET and
    Fundaci{\'o}n Antorchas (Argentina).} 
    and A.~L{\'o}pez\addressmark[CAB]\thanks{Supported by CONICET and 
    ANPCyT through grant No.\ $03-03924$.}}
\begin{document}

\begin{abstract}
  Regularization modifies the (odd) behaviour of the Abelian
  Chern-Simons (CS) action under parity.  This effect happens for any
  sensible regularization; in particular, on the lattice.  However, as
  in the chiral symmetry case, there exist generalized parity
  transformations such that the regularized theory is odd, and the
  corresponding operator verifies a Ginsparg-Wilson (GW) like
  relation. We present a derivation of such a relation and of the
  corresponding symmetry transformations.  \vspace{1pc}
\end{abstract}

\maketitle

It has been shown by L{\"u}scher~\cite{lusch}, that an action defined in
terms of a Dirac operator obeying the GW relation~\cite{gwr} will be
invariant under some symmetry transformations, `deformed' versions of
the usual chiral transformations.  Moreover, those deformed
transformations will tend to the usual ones when the cutoff is removed
(continuum limit).  The existence of this symmetry is more than a mere
curiosity: it has consequences, for example, for the proper
formulation of index theorems and anomalous Jacobians on the
lattice~\cite{Luscher:1999du,Fujikawa:1999ku,Fujikawa:1999if}.  The
symmetry transformations may also be used to understand the anomaly in
a different way: performing a transformation which is a symmetry of
the {\em regularized\/} action, one derives the anomalous conservation
law for a {\em regularized\/} current. The anomalous contribution
comes from the non-trivial Jacobian due to the non-local nature of the
transformations, and requires no additional regularization.  The
survival of a generalized version of the symmetries in a regularized
action (not necessarily on the lattice) is a phenomenon that has also
been explored in other
systems~\cite{Fosco:2001wd,Fosco:2000ud,Ciccolini:2000jc,biet}.

We shall here discuss the case of the CS action, where, as recently
pointed out in~\cite{berr}, the classical transformation properties of
the action are spoiled by the lattice (in fact, by any
regularization). Our aim is to show that there are generalized
transformations in the regularized theory such that the action has the
same parity as the classical one~\cite{Fosco:2001aq}, and also to
present the corresponding GW like relation for the kinetic operator of
the CS action.

To fix our notation and conventions, we shall define the (Euclidean)
CS action $S_{CS}$ by
\begin{equation}
  \label{eq:defscs}
S_{CS} [A]\;=\; i \,\frac{\kappa}{2}\,\int d^3x \, \epsilon_{\mu\lambda\nu} A_\mu \partial_\lambda A_\nu \;,
\end{equation}
where $A_\mu$ is an Abelian gauge field.  $A_\mu(x)$ will be regarded as
the `coordinate representation' of a vector field $|A_\mu \rangle$ in a
Hilbert space, namely, \mbox{$A_\mu (x) \;\equiv\; \langle x | A_\mu \rangle$}.  In terms
of this notation, the CS action is, of course, written as a quadratic
form:
\begin{equation}
  \label{eq:defscs1}
S_{CS} [A]\;=\; i \frac{\kappa}{2}\,\langle A_\mu | R_{\mu\nu} | A_\nu\rangle \;\equiv\; 
i \frac{\kappa}{2} \, \langle A |R| A \rangle \;,
\end{equation}
completely determined by the kinetic operator $R_{\mu\nu} = i \epsilon_{\mu\lambda\nu}
\partial_\lambda$.

Rather than considering the usual parity transformations, we will
follow~\cite{berr}, and use the more symmetric (full) spacetime
inversion $I$:
\begin{equation}
  \label{eq:defprt}
| x \rangle \;\to\; | x^I \rangle \;=\; {\mathcal I} | x \rangle\;\;,\;\; 
\langle x |{\mathcal I} | y \rangle \;=\; \delta (x+y)\;.
\end{equation}
The effect of this transformation on the gauge field
is~\footnote{Since $I$ is just a time reversal times a $\pi$ rotation,
  everything we do for the $I$ transformations has an immediate
  analogy for the parity transformations.}:
\begin{equation}
  \label{eq:defprt1}
\langle x^I | A^I_\mu \rangle = - \langle x | A_\mu \rangle \;\;
{\mathbf I} |A_\mu \rangle =|A_\mu^I\rangle = - {\mathcal I} \, 
|A_\mu \rangle \,,
\end{equation}
where ${\mathbf I} \;=\; - {\mathcal I}$; hence, the odd nature of the
CS action: $S_{CS}[A^I]\;=\; - S_{CS}[A]$, follows from
\begin{equation}
  \label{eq:odd}
{\mathbf I}^t \, R \, {\mathbf I} \;=\; - R \;\;\;
\Leftrightarrow  \;\;\; {\mathbf I} R \,+\, R {\mathbf I} \;=\; 0 \;,
\end{equation}
(we used ${\mathbf I}^t = {\mathbf I}^\dagger = {\mathbf I}$).  As shown
in~\cite{Fosco:2001aq}, the minimal addition one can make to improve
the UV behaviour of the CS propagator, is to include a Maxwell term.
This leads us to the Maxwell Chern Simons (MCS) action~\cite{mcs}
$$
S_{MCS}[A]=\frac{1}{2} \langle A_\mu | \left[ \kappa R_{\mu\nu} - \frac{\kappa^2\partial^2}{M^2}
  \delta^\perp_{\mu\nu}\right] |A_\nu\rangle
$$
\begin{equation}
=\,\frac{\kappa}{2}\, \langle A_\mu | {\tilde R}_{\mu\nu} | A_\nu \rangle
\end{equation}
where $\delta^\perp_{\mu\nu}=\delta_{\mu\nu}- \displaystyle{\frac{\partial_\mu \partial_\nu}{\partial^2}}$ and
${\tilde R}\equiv R (1-\frac{\kappa}{M^2}R)$.  A generalized inversion
${\mathbf{\tilde I}}$ may then be obtained by proposing a general
linear transformation of $|A_\mu\rangle$,
\begin{equation}
   \label{eq:git}
|A^{\tilde I}_\mu\rangle ={\mathbf{\tilde I}} |A_\mu\rangle =- {\mathcal I}\,(f \delta^\perp_{\mu\nu}- g R_{\mu\nu}) |A_\nu\rangle
\end{equation}
with the unknown coefficients $f$ and $g$ to be determined by the
condition:
\begin{equation}
   \label{eq:cond}
{\mathbf{\tilde I}}^t \, {\tilde R}_{\mu\nu} \, {\mathbf{\tilde I}} \;=\; -
{\tilde R}_{\mu\nu} \;.
\end{equation}
(this fixes the transverse part of the transformation; a longitudinal
piece may be added to ${\mathbf{\tilde I}}$ in (\ref{eq:git}) without
affecting (\ref{eq:cond})).  The scalar functions $f$ and $g$ are
found to be:
\begin{equation}
f =\sqrt{1 - \xi^2 \, \frac{\partial^2}{M^2}}
\;,
g=\frac{\kappa}{M^2} \sqrt{1- \xi^2 \,\frac{\partial^2}{M^2}} \;,
   \label{eq:fg}
\end{equation}
with $\xi \equiv \frac{\kappa}{M}$. Choosing for the longitudinal part of the
gauge field the simplest form, we may write
\begin{equation}
   \label{eq:lpt}
{\mathbf{\tilde I}} |A_\mu\rangle \,=\,- {\mathcal I}\,(f \delta_{\mu\nu}\,-\, g R_{\mu\nu})
|A_\nu\rangle \;.
\end{equation}
It is interesting to realize that the generalized inversion defined by
(\ref{eq:git}), may also be written as:
\begin{equation}
   \label{eq:git1}
{\mathbf{\tilde I}}={\mathbf I}
\sqrt{\frac{1-\xi\frac{1}{M} R}{1 + \xi \,\frac{1}{M} R}}
\end{equation}
where the property (\ref{eq:cond}) becomes evident.  The operator
${\mathbf{\tilde I}}$, also shares with ${\mathbf I}$ the property of
being hermitian and idempotent:
\begin{equation}
{\mathbf {\tilde I}}^2 \,=\, {\mathbf I}^2 \,=\,1 \;,
\end{equation}
thus the spectrum of ${\mathbf I}$ is unchanged by our proposed
generalization to ${\mathbf{\tilde I}}$. There are striking
similarities between the properties verified by the regularized
operator ${\tilde R}$ and the Ginsparg-Wilson Dirac operator (in an
even number of dimensions). Indeed, it is just a matter of some
algebra to verify that the operator ${\tilde R}$ fulfills a relation:
\begin{equation}
  \label{eq:gwrp}
{\mathbf I} {\tilde R} + {\tilde R} {\mathbf I} = 2 \frac{\xi}{M} 
{\tilde R} {\mathbf I}(1 - \frac{\xi^2}{M^2} \partial^2)^{-1}  {\tilde R}
\end{equation} 
which looks like a GW relation, if one makes the identifications:
${\mathcal D} \leftrightarrow {\tilde R}$, $\gamma_5 \leftrightarrow {\mathbf I}$, and the
(exponentially local) factor $(1 - \frac{\xi^2}{M^2} \partial^2)^{-1}$ is
interpreted as a diagonal matrix.

Moreover, the hermitian operator ${\hat R} = {\mathbf I} {\tilde R}$
anticommutes with ${\mathbf{\tilde I}}$:
\begin{equation}
  \label{eq:acomm}
{\mathbf{\tilde I}} {\hat R} \,+\, {\hat R} {\mathbf{\tilde I}} = 0\;,
\end{equation}
hence, ${\tilde I}$ may be regarded as the equivalent of the `$\Gamma_5$'
matrix of GW fermions, introduced to study the index theorem for
chiral fermions on the lattice~\cite{Fujikawa:2000bh}.

The introduction of a parity even term is a feature of any sensible
regularization implementable at the Lagrangian level. Indeed, we may
write a general regularized action $S^{reg}$ as
\begin{equation}
S^{reg}[{\mathcal A}] \;=\; S_{PV}[{\mathcal A}] \,+\, S_{PC}[{\mathcal A}]
\end{equation}
where $S_{PV}$ and $S_{PC}$ denote parity violating and conserving
terms, respectively. For example,
\begin{equation}
S^{reg}[{\mathcal A}]= \frac{\kappa}{2} \langle A_\mu | u(\frac{\partial^2}{M^2}) R_{\mu\nu} |A_\nu \rangle 
\nonumber
\end{equation}
\begin{equation}
   \label{eq:greg}
+\,\frac{1}{4} \langle F_{\mu\nu}| v(\frac{\partial^2}{M^2})|F_{\mu\nu}\rangle \;,
\end{equation}
where $u$ and $v$ are chosen in order to obtain the desired behaviour
for the propagator. However, to preserve gauge invariance, we cannot
just set $v\equiv0$, since that would spoil the gauge invariance of the
Chern-Simons action under large gauge
transformations~\cite{Fosco:2001aq}.

Everything we have just mentioned has an immediate analogy in lattice
CS. We first recall a result presented in~\cite{berr}: the most
general form of a local gauge invariant CS action in three dimensions
must include a Maxwell-like term $S_M$ if the action is assumed to be
local, gauge invariant, and free of doublers:
\begin{equation}
   \label{mx}
S_{M}\;=\; \sum_{x,y} A_\mu (x) M_{\mu\nu}(x-y) A_\nu (y)
\end{equation}
where
\begin{equation}
   \label{maxwell}
 M_{\mu\nu}(x-y)=- \Box  \, \delta_{\mu\nu} + d_\mu {\hat d}_\nu \;,
\end{equation}
with $ \Box = \sum_{\mu =0}^2 d_\mu {\hat d}_\mu$ the lattice Laplacian in three
dimensions (see \cite{berr} for details).

As in the continuum, parity and inversion are violated. However, as we
have shown for the continuum case, a generalized version of that
transformation survives.  On the lattice, they are more conveniently
formulated in the Fourier representation: with the lattice Fourier
transformation of the gauge field $A_{\mu}$ defined by
\begin{equation}
   \label{ftrans}
 A_\mu (x) = \int_{\cal B} {\frac {d^3p}{(2\pi)^3}} e^{-i px} e^{-i {\frac{p_{\mu}}{2}}}
{\tilde  A}_{\mu}(p) \;,
\end{equation}
where the integration region is the first Brillouin zone ${\cal B}$,
the Fourier space representation of the Chern-Simons action becomes
\begin{equation}
   \label{lcsft}
S_{CS}\;=\; \int_{\cal B} {\frac {d^3p}{(2\pi)^3}} {\tilde  A}_{\mu}(p)
{\tilde  G_{\mu\nu}}(p) {\tilde  A}_{\nu}(-p)
\end{equation}
with ${\tilde G_{\mu\nu}}(p)= e^{-i {\frac{p_{\mu}}{2}}} G_{\mu\nu}(p)
e^{i{\frac{p_{\nu}}{2}}}$.

The lattice MCS action can be written as
\begin{equation}
   \label{mcsft}
S\;=\; \int_{\cal B} {\frac {d^3p}{(2\pi)^3}} {\tilde  A}_{\mu}(p)
\Gamma_{\mu\nu}(p) {\tilde  A}_{\nu}(-p)
\end{equation}
where $\Gamma_{\mu\nu}(p) = f(p) \delta_{\mu\nu}^{\perp} (p) + i g(p) Q_{\mu\nu}(p)$, with
\begin{equation}
f(p) = {\frac {4}{e^2}} {\sum_{\alpha=0}^{2} \sin^2{\frac {p_\alpha}{2}}} \;,
\end{equation}
\begin{equation}
g(p) = 2 h(p)  \sqrt{{\sum_{\alpha=0}^{2} \sin^2{\frac {p_\alpha}{2}}}}  
\end{equation}
and
\begin{equation}
\delta_{\mu\nu}^{\perp} = \delta_{\mu\nu} - {\frac {\sin({\frac {p_\mu}{2}})
\sin({\frac {p_\nu}{2}})}{{\sum_{\alpha=0}^{2} \sin^2({\frac {p_\alpha}{2}})}}} \,,
Q_{\mu\nu} =  \epsilon_{\mu\alpha\nu} \frac{{\hat p}_\alpha}{\sqrt{{\hat p}^2}}\,.
\end{equation}
We obtain the generalized parity transformations following the same
steps as in the continuum case.  In the Fourier representation, the
gauge field transformation is given by
$$
A_{\mu}^I(p^I)= i {\frac {f(p)}{\sqrt {f^2(p)+g^2(p)}}} Q_{\mu\rho}
A_{\rho}(p)
$$
\begin{equation}
-{\frac {g(p)}{\sqrt {f^2(p)+g^2(p)}}}\; \delta_{\mu\rho}^{\perp} A_{\rho}(p)
\end{equation}
that tends to the usual inversion when $f \to 0$. Therefore , this is
the new parity transformation under which the Chern-Simons action is
still odd but the kernel is integrable, and whose existence was
suggested in reference~\cite{berr}.

Thus we conclude that in $2+1$ dimensions there exists an identity,
analogous to GW relation, involving the inversion (or parity) operator
and the kinetic operator which defines the CS action.  Also, this
relation allows for the definition of a generalized inversion
operator, under which the regularized action has the same properties
as the unregularized one.


\begin{thebibliography}{9}
\bibitem{lusch}M.~L{\"u}scher, Phys.\ Lett.\ {\bf B428}, 342 (1998).
\bibitem{gwr}P.~Ginsparg and K.~Wilson, Phys.~Rev.~{\bf D25}, 2649
  (1982).  
\bibitem{Luscher:1999du} M.~L{\"u}scher, Nucl.\ Phys.\ {\bf B549}, 295 
(1999).  
\bibitem{Fujikawa:1999ku} K.~Fujikawa, Phys.\ Rev.\ {\bf D60}, 074505 
(1999).
\bibitem{Fujikawa:1999if}K.~Fujikawa, Nucl.\ Phys.\ {\bf B546}, 480
  (1999).  
\bibitem{Fosco:2001wd} C.~D.~Fosco and M.~Teper, 
Nucl.\ Phys.\ {\bf B597}, 475 (2001).  
\bibitem{Fosco:2000ud}C.~D.~Fosco and F.~D.~Mazzitelli, 
Phys.\ Lett.\ {\bf B481}, 129 (2000).
\bibitem{Ciccolini:2000jc}M.~L.~Ciccolini, C.~D.~Fosco and
  F.~A.~Schaposnik, Phys.\ Lett.\ {\bf B492}, 214 (2000).
\bibitem{biet}W.~Bietenholz and J.~Nishimura, {\em `Ginsparg-Wilson
    fermions in odd dimensions'}, hep-lat/0012020.
\bibitem{berr}F.~Berruto,~M.~C.~Diamantini and P.~Sodano, Phys.\ 
  Lett.\ {\bf B487}, 366 (2000).  
\bibitem{Fosco:2001aq} C.~D.~Fosco and A.~L{\'o}pez, Phys.\ Rev.\ 
{\bf D64}, 025017 (2001).
\bibitem{mcs}S.~Deser, R.~Jackiw and S.~Templeton, Ann.~of
  Phys.~(N.Y.) {\bf 140}, 372 (1982).
\bibitem{Fujikawa:2000bh}K.~Fujikawa, Chin.\ J.\ Phys.\ {\bf 38}, 551
  (2000).
\end{thebibliography}
\end{document}